# Investigating the Permeability Evolution of Artificial Rock During Ductile and Brittle Deformation Under Pressurized Flow

Shaimaa Sulieman[1], Martin Stolar[1], Ludmila Abezgauz[1], Shouceng Tian[2], and Yaniv Edery[1]

[1]Faculty of Civil and Environmental Engineering, Technion, Haifa, Israel.

[2]State Key Laboratory of Petroleum Resources and Prospecting, China University of Petroleum (Beijing), Beijing, 102249, China

Abstract

The drilling of geothermal energy, $CO_2$ sequestration, and wastewater injection all involve the pressurized flow of fluids through porous rock, which can cause deformation and fracture of the material. Despite the widespread use of these industrial methods, there is a lack of experimental data on the connection between the pore pressure rise, the deformation and permeability changes in real rock. In order to address this gap in the literature, this study developed an artificial rock material that can be deformed and fractured at low pressures. By controlling the porosity, permeability, and strength of the material during the sintering process, it is possible to mimic various types of rock. The artificial rock was designed to accommodate radial flow and deformation, allowing for the tracking of deformation by monitoring the flux and driving pressure and thus calculating the permeability changes under various pressure conditions. The study was able to examine the impact of both ductile and brittle deformation on the permeability during pressurized flow, which were captured by two models that were adjusted to this scenario. This study provides a link between pressurized flow, rock formation permeability and ductile to brittle deformation, that can constrain risk assessment to geothermal energy and $CO_2$ sequestration.

1. Introduction

The interplay between fluid injection and the mechanical response of granular materials is a subject of paramount importance in various geomechanical applications, including hydraulic fracturing, underground gas storage, and groundwater management [1–5]. This interaction is particularly significant in porous rocks, where fluid injection leads to the formation and propagation of fractures, enhancing the permeability of the reservoir and facilitating hydrocarbon extraction [6–10]. The behavior of porous media under fluid injection is complex and governed by various physical processes, including changes in porosity and permeability that affect fluid flow [11–14]. Additionally, the interaction between the fluid and solid matrix can induce shear banding, compaction, and failure, depending on rock properties and injection conditions [15–19].

With the growing need for underground storage of hydrogen and gas due to the intermittent nature of renewable energy, permeable rock formations like sandstones have become essential for rapid injection and withdrawal of natural gas or hydrogen [20–22]. However, the ductile and brittle failure of the rock formation leads to sand production, which is a persistent challenge in sandstone reservoir operations, affecting well productivity and storage performance [23–25]. Under hydrostatic loading, this deformation causes permeability and porosity decrease with increasing confining pressure due to the closure of pores and micro-cracks [26–31]. As we will show here, even for high porosity samples permeability may still decrease even in the dilatant stage, possibly due to sand production and increased tortuosity, during loading, necessitating continuous measurements of gas permeability and porosity under deviatoric stress [17,32,33].

Recent studies have employed numerical simulations, laboratory experiments, and field observations to investigate fluid-driven deformation of porous media [1,2,34–38]. Despite advances, challenges remain in accurately predicting and controlling fluid-driven deformations due to the heterogeneity of natural rocks, complexity of governing equations, and uncertainties in material properties [33,39–41]. Further research is required for sustainable and responsible utilization of subsurface resources [42–45].

In this study, the coupling between permeability change and deformation in PMMA rock-like porous material was systematically investigated, focusing on the failure of the sample and the transition between ductile and brittle deformation. Experimentally, incremental pressure differences across various injection cycles were applied, enabling the distinction between ductile and brittle regimes. The material's response was monitored through permeability calculations, revealing a non-uniform reduction in permeability during the ductile regime and abrupt increases during the brittle regime. The modeling involved the derivation of functional forms from pulling tests for the elastic (linear) and elasto-plastic (power-law) material responses, using stress-strain curves. The threshold effective stress marking the transition from elastic to plastic deformation was identified. The elastic or elasto-plastic poro-mechanic equation was solved for radial and angular components, enabling the determination of radial and angular strains, which were then related to porosity changes. The permeability change was calculated for each local porosity change and averaged radially to ascertain the total permeability change due to pressurized flow. This comprehensive approach to understanding the complex interactions between fluid injection and porous media deformation contributes valuable insights to the field of geomechanics, with potential applications in energy storage, resource extraction, and environmental management.

2. Methods

2.1 Experimental setup

The experimental setup employed a Fluigent Flow Ez 7000 system to precisely regulate inlet pressure, in conjunction with meticulous monitoring of fluid flow through the Fluigent Flow unit L at 100 ms intervals, utilizing Fluigent A-IO software. Calibration of both the flow unit and pressure controller was achieved for various flow rates and induced pressures by assessing the fluid weight at the outlet to calculate volume, and by gauging the pressure at the inlet with a pressure sensor (refer to figure 1 for experimental layout, and figure S1 a and b in the supplementary material for calibration curves). The pressure differential is generated by modulating the fluid's inlet pressure, allowing unobstructed radial flow towards an outlet at the sample's periphery, exposed to ambient atmospheric pressure, a condition akin to pressurized well flow. Following solidification and drying, the sample was saturated with a thoroughly characterized vegetable oil, optically congruent with the refractive index of the beads, as ascertained by a refractometer (Schmidt and Haensch ABBE AR12.), thus enabling observational studies.

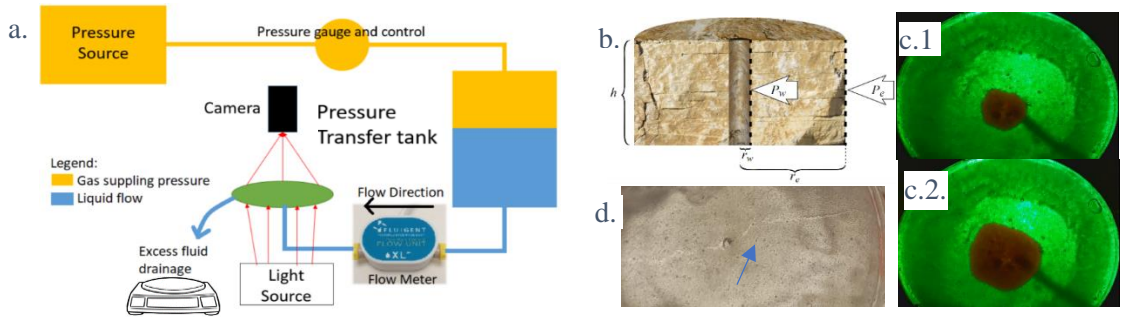

Figure 1a. An illustration of the experimental setup. b. An illustration of the sample radial flow condition and dimensions. c. A radial invasion of a dyed fluid for 2 different times (c.1. and c.2.). d. brittle fracture in one of the samples.

2.2 Material mechanical properties

Poly-Methyl-Methacrylate (PMMA) beads were sintered to fabricate a circular sample, situated within a PMMA petri dish, employing an acetone and ethanol solvent mixture with varying ratios (3%-20%). This solvent mixture functioned to dissolve the outer layer of the beads, which, upon evaporation under controlled vacuum conditions, re-solidified, interlocking with adjacent beads, thereby enabling control over the sample strength. The size of the beads was ascertained using a MasterSizer 3000 laser-based diameter measuring instrument, revealing a size range of 50 to 200 microns, with a mean value of 130 microns (refer to figure S2 in the supplementary material for additional details). Both pre- and post-sintering processes, the bead weight and height were determined. The acetone concentration varied from $C_{aceton} = 3$ to 20%, with concentrations below 3% failing to form a stable skeleton, and concentrations above 20% leading to complete dissolution of the beads.

To investigate the properties of the sample material and their influence on deformation behavior, a pulling test experiment was conducted to measure the Young's modulus, plastic modulus, and tensile strength (figure 2). These parameters were subsequently scaled with the measured porosity for each solvent concentration. The pulling test involved a rectangular specimen subjected to a direct Young's modulus test, followed by a plastic modulus regime, culminating in a tensile strength test where the sample fully yielded, employing a method known as the dog bone test. This test was performed on a rectangular specimen with dimensions of $6.5 \times 1.5 \times 0.2\ [cm]$. The ends of this rectangle were immersed in slow-hardening epoxy to mitigate damage by the clamps of the material testing machine (Lloyd LF-Plus with XLC Series

50N load cell). These clamps exerted equal but opposing forces on the sample until fracture. The total stress ($S_{tot}$) of the sample was computed using the following formula:

1. $$S_{tot} = F/A$$

where $F$ is force applied at fracture and $A$ is cross sectional area of the sample tested.

In figure 1, the stress ($S$) to strain ($\varepsilon$) curve for three distinct samples is depicted, and the Young's modulus is measured directly from the slope of the elastic expansion, given by $S_e = E\varepsilon_e + 763\ [Pa]$, and $E = 1.712 \cdot 10^7\ [Pa]$. Utilizing previous work in which the Odometric modulus ($E_{oed}$) for the elastic expansion of the sample was identified [11], the Poisson's ratio ($\nu$) for the entire sample is calculated as follows:

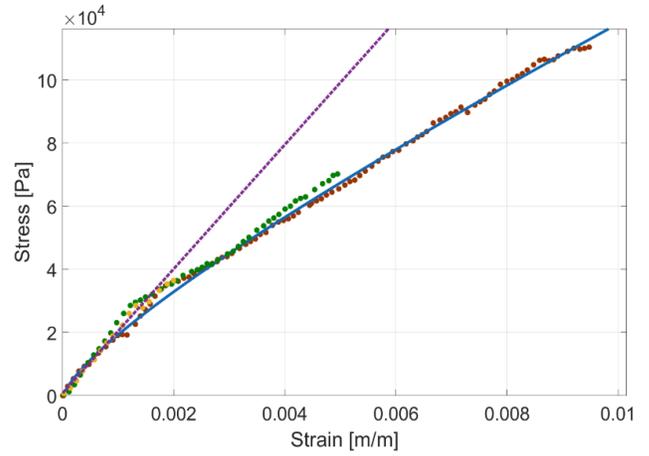

Figure 2. Stress-strain curve of the artificial rock obtained through a pulling test. The filled circles denote three distinct samples. The dotted line represents a linear fit for the elastic response, given by $S_e = 1.72 \cdot 10^7 \varepsilon_e + 763$, with an $R^2 = 0.987$, where the slope designates the Young's modulus of the sample. The solid line corresponds to a power-law fit to the elasto-plastic response of the sample, expressed as $S_p = 9.3 \cdot 10^5 \varepsilon_p^{0.63} + 6.37 \cdot 10^6 \varepsilon_p - 1265$, with an $R^2 = 0.998$. The cut-off for each experiment, where the sample yielded, is demarcated by the termination of the dotted line, and its variability is attributed to torque forces on the sample, leading to fluctuations in yielding. Nonetheless, it is evident that the sample's transition from elastic to elasto-plastic response occur at a strain of $\varepsilon_p \approx 0.001$, and stress of $S_p \approx 2 \times 10^4\ Pa$.

2. $E_{oed} = 1.78 \cdot 10^8\ [Pa] = \dfrac{E}{1 - \dfrac{2\nu^2}{1-\nu}} \to \nu = 0.48$

Given that the PMMA Poisson ratio for the bulk material is $\nu_{bead} = 0.37$ (refer to table 1 in Alam et al., (2017) [46]), the Poisson ratio for the porous material as a whole is found to be within acceptable bounds. Furthermore, we recognize the alteration in the slope following the elastic response, indicative of the elasto-plastic transition preceding the yielding of the sample. This elasto-plastic response is duly accounted for, and the plastic strain, $\varepsilon_p$, is extracted through a power law fit, described as follows:

3. $\quad S_p = E_{ep}^e \varepsilon_p^\beta + E_{ep}^p \varepsilon_p - 1265$

where the plastic response is characterized by $E_{ep}^p = 9.3 \cdot 10^5$, and $\beta = 0.63$, while the elasto-plastic response is denoted by $E_{ep}^e = 6.37 \cdot 10^6$. The fitting parameters obtained for both the elastic and elasto-plastic curves will subsequently be employed to compute the local strain attributable to the radial pressure drop in the flow experiment, in accordance with the law of effective stresses.

To ascertain the porosity, the weight of both the saturated and unsaturated samples is measured to determine the pore volume. This method has been corroborated by a CT scan of the sample, demonstrating its efficacy for our particular samples (refer to [11] for details). Utilizing the porosity, the initial permeability, denoted as $k_{in}$, is computed employing the Kozeny-Carman relation for spherical beads, described as follows:

4. $$k_{in} = \frac{\varphi_{in}^3}{180 d^2 \psi^{-1}(1-\varphi_{in})^2}$$

where $\varphi$ represents the porosity, $d$ is the mean bead size, and by assuming random packing, the parameter $\psi$ is set to one.

2.3 Pressurized flow equations

2.3.1 Effective stress for the elastic case

With knowledge of the fluid properties, specifically the kinetic viscosity ($\mu_k = 62.2\ [cp]$), the permeability measurement was corroborated by the experimental setup. This was achieved by applying a pressure drop and measuring the flux, in accordance with the Darcy equation expressed in polar coordinates, as follows:

5. $$q = \frac{2\pi h k_{in}(P_e - P_w)}{\mu_k \ln\left(\frac{r_w}{r_e}\right)}$$

where $q$ is the flow rate, $h$ is the height of the porous medium, $P_w$ is the applied inlet pressure at $r_w$, the inner radii of the sample, and $P_e$ is the outlet pressure at $r_e$, the outer radii of the sample, set to atmospheric pressure (see figure 1.b. for dimensions and figure 1.c.1 and 2 for the radial invasion). Assuming that in the elastic case there are small perturbation to the porosity [47], the minor changes in strain will not affect the overall permeability of the sample, denoted as $k_{in}$. We follow the equilibrium equation for the effective stress, in accordance with the Cauchy total stress tensor for axisymmetric deformation [48]:

6. $$\frac{\partial s_r}{\partial r} + \frac{dP}{dr} = \frac{s_\alpha - s_r}{r}.$$

For higher deformations, the role of effective stress in poroelasticity stipulates that the total stress for each location $S_r$ is equal to the sum of the local pore pressure, $P_r$ multiplied by the Biot coefficient $\theta$, and the Tarzaghi's effective stress tensor $S_{tar}$. For axisymmetric deformation in a plane-strain condition, this relationship yields the following equation in radial coordinates:

7.
$$S_r = S_{tar}^r - P_r \cdot \theta$$
$$S_\alpha = S_{tar}^\alpha - P_r \cdot \theta$$

The Biot coefficient is computed from the bulk moduli ($E$) for the porous skeleton and the solid material moduli ($E_s = 2.8 \times 10^9 \ Pa$), in accordance with the methodology described by Selvadurai & Suvorov [49]:

8.
$$\theta = 0.994 = 1 - \frac{E}{E_s}$$

and the $P_r$ is directly derived from equation (5), by knowing the corresponding $P_e$ and $P_w$ for $r = r_e$, and $r = r_w$, respectively, while monitoring $q$ during the experiment. It is further assumed that the stress and strain along the z-axis are negligible, owing to the small cross-section of the sample in comparison to its radii. We then adhere to the convention that the radial stress at the inlet is equal to the applied pressure, as follows:

9.
$$S_r(r_w) \equiv -P_w$$

as the outlet pressure is $P_e(r_e) = 0$. However, owing to the radial nature of the sample, the stress is expected to diminish as the square of the radii [48,50], and considering that the pressure at the inlet is the sole contributor to the stress, we receive the following expression:

10.
$$S_r(r) \equiv -P_w \cdot \frac{r_w^2}{r^2},$$

and the effective stress in the radial direction takes the form of:

11.
$$S_{eff}^r(r) = -P_w \cdot \frac{r_w^2}{r^2} - P_r \cdot \theta$$
$$S_{eff}^\alpha(r) = P_w \cdot \frac{r_w^2}{r^2} - P_r \cdot \theta.$$

This elementary derivation for the effective stresses, which will be refer to from here on as the simplistic calculation, results from the Biot coefficient, which is proximate to one, and the elevated Poisson ratio. As we have no horizontal in-situ stress in a normally stressed region, and our stress is lined up with the principal radial stress, there is no need to rotate the coordinate frame with the in-situ components as in Bradley (1979) [50]. Moreover, these factors render the additional terms derived Wang (1996) [48] ($\beta$ and the $\lambda$ and $G$ Lame's elastic coefficients parameter) negligible. These terms are calculated to be two orders of magnitude smaller due to the high porosity in our system and the stiffness of the material. We also provide a more thorough calculation for the effective stresses, which will be refer to from here on as the fuller calculation, following Auton and MacMinn [2] in the following sections.

2.3.2 Strain for the elastic case

In accordance with Coussy (2004) [47], we define $S_{tar}$ using the isotropic linear theory, where the plane strain condition for the elastic component necessitates an axisymmetric deformation for radial coordinates. Consequently, $S_{tar}(r)$ and $S_{tar}(\alpha)$ can be expressed in the following form:

12.1. $$S_{tar}^e(r) = C_1 \cdot (\epsilon_r^e + \epsilon_\alpha^e) + C_2 \cdot \epsilon_r^e$$

12.2. $$S_{tar}^e(\alpha) = C_1 \cdot (\epsilon_r^e + \epsilon_\alpha^e) + C_2 \cdot \epsilon_\alpha^e$$

Given that $C_1, C_2, \epsilon_r$, and $\epsilon_\alpha$ are defined as follows: $C_1 = \frac{3K\nu}{(1+\nu)}$, $C_2 = \frac{3K(1-2\nu)}{(1+\nu)}$, $\epsilon_r^e = \varepsilon_e$, $\epsilon_\alpha^e = \frac{\epsilon_r^e}{r}$, where $\nu$ is calculated from the Odometric test, from which the bulk modulus is calculated $K = \frac{E}{3(1-2\nu)}$, as shown by [51].

In our experiments we monitor the pressure and flow rate for 20 seconds to attain steady-state conditions wherein the measured pressure, flow rate, and consequently the permeability remain constant. As we will subsequently demonstrate, our sample is primarily ductile in nature; thus, it will plastically deform prior to undergoing substantial elastic deformation, which is presumed to be quasi-static and reversible. Accordingly, utilizing equation (6) and the boundary condition stipulating that $S_{eff}^r(r_w) = S_{eff}^\alpha(r_w)$, in conjunction with equations (11) and (7), and equations (12.1) and (12.2) the following expression:

13. $$S_{eff}^r(r)_e = C_1 \cdot (\epsilon_r^e + \epsilon_\alpha^e) + C_2 \cdot \epsilon_r^e$$
$$S_{eff}^\alpha(r)_e = C_1 \cdot (\epsilon_r^e + \epsilon_\alpha^e) - C_2 \cdot \epsilon_\alpha^e$$

Equation (13) can be directly solved utilizing the parameters measured in our experimental setup, specifically, $P_w$, $q$, $k_{in}$, $E$, and $\nu$, in conjunction with the boundary conditions for the inlet and outlet pressure.

2.3.3 Effective stress for the elasto-plastic case

For the plastic case we adhere to the Mohr-Coulomb condition for cohesive-frictional yield condition given by $|\tau'| \leq -S_{eff}\tan(\omega) + c$, as illustrated in [52]. Here $\omega$ represents friction angle taken to be 36°, since our sample exhibits characteristics akin to sandstone with moderate internal friction and cohesion [53–55]. Following the derivation for the elasto-plastic case as presented in Auton & MacMinn [2], we calculate the following: $W = \frac{1+\sin(\omega)}{1-\sin(\omega)}$, and $y = \frac{2S_p(\varphi)\cos(\omega)}{1-\sin(\omega)}$, where $S_p(\varphi)$, the tensile strength as measured by the pulling test, will be defined in section 2.3.5. These parameters enable us to delineate the transition between the plastic and elastic radial regions within the sample:

14. $$W \cdot S_{eff}^{\alpha}(r)_e' = y + S_{eff}^{r}(r)_e'$$

We further define the dilation angle $\Psi = 2°$, analogous to sandstone [54], yielding $\beta = \frac{1+\sin(\Psi)}{1-\sin(\Psi)} = 1.072$. hese parameters facilitate the definition of equations (26) and (29) in Auton & MacMinn [2], namely:

15.1. $$S_{eff}^{r}(r)_e = \frac{S_r(r_w)}{r \ln(\frac{r_e}{r_w})}, \quad S_{eff}^{\alpha}(r)_e = \frac{y + S_{eff}^{r}(r)_e}{W}, \quad r_{e\ to\ ep} < r < r_e$$

15.2. $$S_{eff}^{r}(r)_p = C_{A1} + C_{A2} \cdot \left(\frac{r}{r_w}\right)^{\beta}, \quad S_{eff}^{\alpha}(r)_p = \frac{y + S_{eff}^{r}(r)_p}{W}, \quad r_w < r < r_{e\ to\ ep}$$

Where $C_{A1} = \frac{y \ln(\frac{r_e}{r_w}) + W S_r(r_w)}{(W-1)\ln(\frac{r_e}{r_w})}$, and $C_{A2} = -C_{A1}$, in our fully permeable case.

2.3.4 Strain for the plastic case

As the elastic strain is predicated on the coefficients measured by the pulling test, equation (13) can be further extended to the plastic case, drawing upon the fitting parameters for the elasto-plastic case as depicted in equation (2). Consequently, equations 12.1 and 12.2 are transformed into the following form:

16.1 $$S_{eff}^{r}(r)_p = C_3 \cdot (\epsilon_r^p + \epsilon_\alpha^p) + C_4 \cdot \epsilon_r^p + C_5 \cdot ((\epsilon_r^p)^\beta + (\epsilon_\alpha^p)^\beta) + C_6 \cdot (\epsilon_r^p)^\beta$$

16.2 $$S_{eff}^{\alpha}(r)_p = C_3 \cdot (\epsilon_r^p + \epsilon_\alpha^p) + C_4 \cdot \epsilon_\alpha^p + C_5 \cdot ((\epsilon_r^p)^\beta + (\epsilon_\alpha^p)^\beta) + C_6 \cdot (\epsilon_\alpha^p)^\beta$$

Given that the elasto-plastic component adheres to a similar parametrization as the elastic part, it can be expressed as follows: $C_3 = \frac{3E_{ep}^e \nu}{(1+\nu)}$, and $C_4 = \frac{3E_{ep}^e(1-2\nu)}{(1+\nu)}$, while the plastic response exhibits a power law scaling with a similar plastic parameterization, specifically expressed as: $C_5 = \frac{3E_{ep}^p \nu}{(1+\nu)}$, $C_6 = \frac{3E_{ep}^p(1-2\nu)}{(1+\nu)}$.

To transition from the elastic response to the elasto-plastic response, we identify the value at which the elastic fit diverges from the plastic fit ($S_{\text{e to ep}} = 2 \times 10^4 \, Pa$) as previously depicted in figure 2. We then switch the calculation of the total strain component from the elastic ($\epsilon_r^e, \epsilon_\alpha^e$) to the elasto-plastic ($\epsilon_r^p, \epsilon_\alpha^p$) when $P_w \cdot \frac{r_w^2}{r^2} - P_r \cdot \theta > S_{\text{e to ep}}$ for the simplistic case. Similarly, we transition between equations (15.1) and (15.2), but we identify the transition from the plastic regime to the elastic regime according to the radii ($r_{\text{e to ep}}$) found at the cusp of the transition (see figure 3). We then calculate the roots of equations (16.1) and (16.2) iteratively, based on the measured parameters in our setup, specifically, $P_w$, $q$, $k_{in}$, $E_{ep}^e$, $E_{ep}^p$, and $\nu$, in conjunction with the boundary conditions for the inlet and outlet pressure.

### 2.3.5 Extending the transition stress from elastic to plastic for various cases.

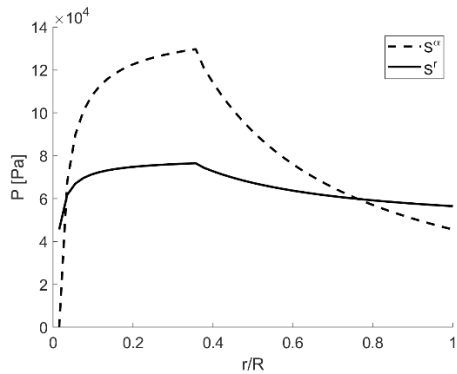

Figure 3. Calculated stress vs normalized radii for a sample with a given inlet pressure. The rise for the radial and tangential stress marks the plastic regime, while the rapid reduction marks the elastic regime, while the cusp marks the radial location at which the transition occurs.

As we modify the preparation stage to encompass a range of solvent concentrations, we account for the non-monotonic change in porosity, specifically, $C_{aceton} = 5, 7.5, 10, 12.5,$ and $15\%$, corresponding to $\varphi_{in} = 0.31, 0.33, 0.36, 0.32,$ and $0.225$, respectively. Given that the variations in porosity are minimal, we can postulate that both the elastic and plastic component variations are negligible. However, the transition between the elastic and plastic response is correlated with the overlapping of the beads, which is a function of the porosity, assuming that their sizes remain constant. Consequently, we employ Balshin's formula [56] to relate the porosity to the transition from the elastic to the elasto-plastic response:

17. $$S_p(\varphi) = S_0(1-\varphi)^b$$

here $S_p(\varphi) \approx 2 \times 10^4 \, Pa$ is the transition between elastic to elasto-plastic measured by the pulling test for $\varphi = 0.36$, $\sigma_0$ is tensile strength of material at zero porosity. For PMMA, this value is 42 MPa, leading to the calculation of the power exponent that relates the tensile strength to the porosity as $b \cong 21$. Consequently, we utilize equation (17) to update the transition value for the total effective stress, denoted as, $S_{\text{e to ep}}$, for each total stress in our experiment at each radius ($S_{eff}^r(r)_e$, and $S_{eff}^\alpha(r)_p$).

We replicate the same process for the Ryshkewithch model, as in certain instances, it may be more appropriate for materials with high porosity [57,58]:

18. $$S_p(\varphi) = S_0 e^{-n\varphi},$$

and then derive the value of n from the measured case and utilize it to calculate $S_{\text{e to ep}}$ or all concentrations. Upon comparison, we did not identify any significant differences between the models, and therefore, we present only the results based on the Ryshkewithch model.

2.3.6 Calculating the permeability from the strain

Utilizing the calculated strain from the local stress, we update the local porosity value by multiplying the initial porosity with the total strain $(1 - \epsilon_{tot}(r))$, which is considered a volumetric strain. However, while for the elastic case we can assume that $\epsilon_r^e$, and $\epsilon_\alpha^e$ a are additive, for the elasto-plastic case, this should be contingent on the type of deformation, whether it is ductile or brittle in nature.

For the ductile case, both the radial and angular components lead to compaction, and therefore they are additive. Conversely, in the brittle case, the opening of a fracture in a specific radial location, marked by the angular strain, contradicts the radial compaction at that location, and it is therefore subtracted. This distinction between the ductile and brittle deformation yields the following criteria:

19. $$\epsilon_{tot}^e(r) = \epsilon_r^e(r) + \epsilon_\alpha^e(r), \, \epsilon_{tot}^p(r) = \epsilon_r^p(r) \pm \epsilon_\alpha^p(r)$$

We employ both criteria for the plastic deformation and multiply the total strain with the initial measured porosity, presuming that the volumetric strain corresponds to the total strain. Utilizing equation (4), we can update the permeability for each radius and for each applied pressure, expressed as follows:

20. $$k_r(P) = k_{in} \frac{(\varphi_{in} \cdot (1-\epsilon_{tot}(r)))^3}{\left(1-\varphi_{in} \cdot (1-\epsilon_{tot}(r))\right)^2} \cdot \frac{(1-\varphi_{in})^2}{\varphi_{in}^3}$$

To obtain the permeability of the entire flow cell, we execute a parallel radial flow-weighted average over the radial change in permeability, articulated as follows:

21. $$k(p) = \frac{ln\left(\frac{r_e}{r_w}\right)}{\sum_i \frac{ln\left(\frac{r_i}{r_{i-1}}\right)}{k_{r_i}(P)}}$$

Alternatively, the change in permeability can also be directly related to the porosity following the power law relation [5,30,59].

22. $$k = k_{in} \left(\frac{\varphi}{\varphi_{in}}\right)^\alpha$$

Where $\alpha$ is the porosity sensitivity exponent, a material constant typically within the range of 2–3 [60,61]. For this study, we selected $\alpha = 3$. We then compare the change in permeability, calculated from the strain effect on the porosity for each applied pressure, with the measured permeability. The latter is directly calculated from equation (5), as we monitored the applied pressure and volumetric flow rate within our system.

3. Results

In the present investigation, the focus was directed towards understanding the interplay between changes in permeability and deformation within a PMMA rock-like porous medium, with particular emphasis on the transition from ductile to brittle deformation. It is essential to recognize that natural rocks exhibit a spectrum of elastic and plastic characteristics, along with variations in porosity and permeability. The methodology employed facilitated the classification of samples into six distinct groups, with four groups yielding results of substantive relevance. These classifications were primarily determined by the acetone concentration in the solvent mixture, which governed the extent of the outer layer dissolution of the PMMA beads, and consequently, their fusion with one another.

The PMMA rock-like material's properties, including grain size and shape, porosity, and material strength, were compared to a broad array of rocks, with particular attention to sedimentary rocks of significant importance for hydraulic deformation. The comparison revealed that the PMMA samples closely resembled the properties of sandstone. However, a notable distinction between the PMMA rock-like material and natural rock lies in the shape of the consolidated grains. The PMMA grains are highly spherical, maintaining a consistent surface-to-radius ratio, a condition seldom encountered in nature. This spherical form results in an idealized porous material structure known as rhombohedral packing, given a narrow radius distribution for the grains. This structure would markedly differ from that of natural rocks. Nevertheless, the utilization of a wide range of grain sizes in the rock-like material for this study ensured a heterogeneous bead packing that deviated significantly from rhombohedral (refer to the supplementary figure S2). Consequently, the synthesized rock-like material can be considered a suitable model for simulating processes occurring in various sandstone sedimentary rocks.

3.1. Identifying brittle deformation

The properties of the beads constituting the sample are designed to emulate rock characteristics, yet the experimental boundary conditions diverge substantially from conventional core tests conducted on actual rock. Standard core tests typically employ a one-dimensional permeability assessment, with the core being jacketed, thereby facilitating a confining stress either aligned with the flow direction [5,17,62,63] or applied triaxially to the entire sample [32,35]. Conversely, most drilling wells induce radial flow over a restricted cross-section, where the inlet pressure propelling the flow [2,64,65] also serves as the stressing force on the boundary. This contrasts with the standard 1D rock core tests.

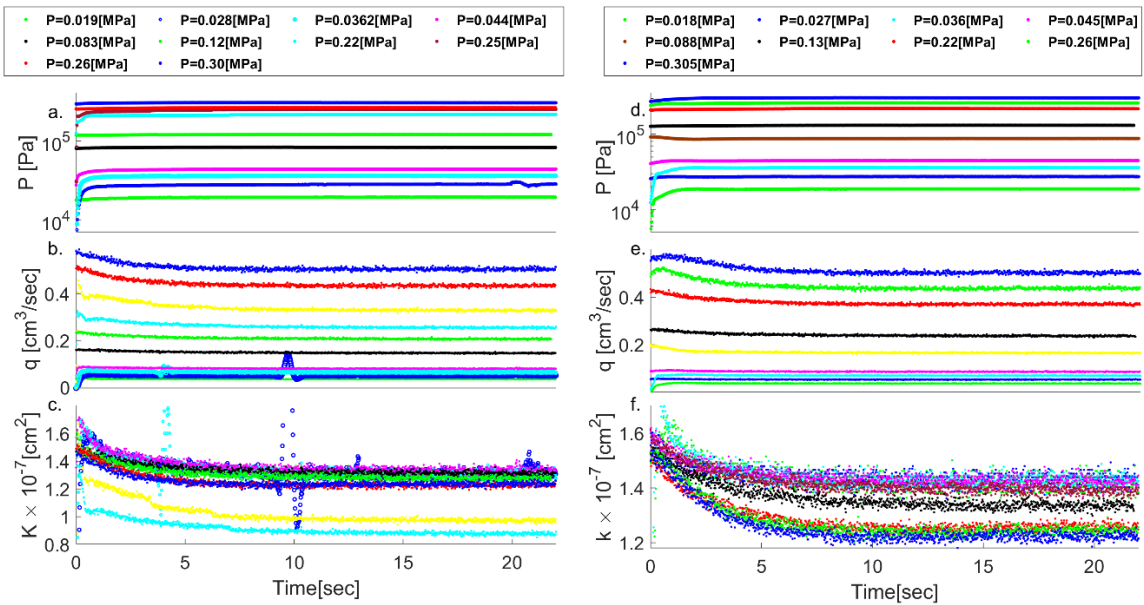

Figure 4. Measurements of a. the applied pressure at the inlet, b. the volumetric flux, and c. the permeability calculation from the measured pressure and flux, for a radial flow experiment under increasing applied pressure at the sample center. Repetition of the same flow experiment, on the same sample, for non-ordered applied pressure, while measuring d. the applied pressure at the inlet, e. the volumetric flux, and f. calculating the permeability.

The sample in this study can be sintered to simulate various geometries and boundary conditions, thereby reproducing saturated pressurized flow applied radially from a circular inlet akin to well injection. Each inlet pressure is maintained for 15-35 seconds to ascertain that the flow, in conjunction with the elasto-plastic deformation, attains a steady state, without further alterations in the measured flux and pressure (refer to figure 4a and b). As depicted in figure 4a, each applied pressure triggers a pressure build-up in the sample (as gauged by the inlet pressure transducer), inducing a corresponding flux increase (figure 4b) for the initial 0.5~1 second. This is followed by a slower pressure reduction, culminating in an asymptotic value, while the flux gradually diminishes over the subsequent 2-4 seconds.

This flux and pressure alteration, attributable to a uniform radial compressive effective stress, has been previously demonstrated in 2D modeling [2,64], and in triaxial compression for 1D core samples [17,35]. The compression may be classified as elastic or as a ductile, plastic deformation. However, there are also sudden flux surges, exceeding the elastic modulus pressure limit, which can solely be ascribed to brittle fractures in the sample, as the permeability escalates in response to them (indicated by circles in figure 4b and c). These brittle fractures are also discernible in the sample, considering that they are continuous and not localized micro-fractures (see figure 1d).

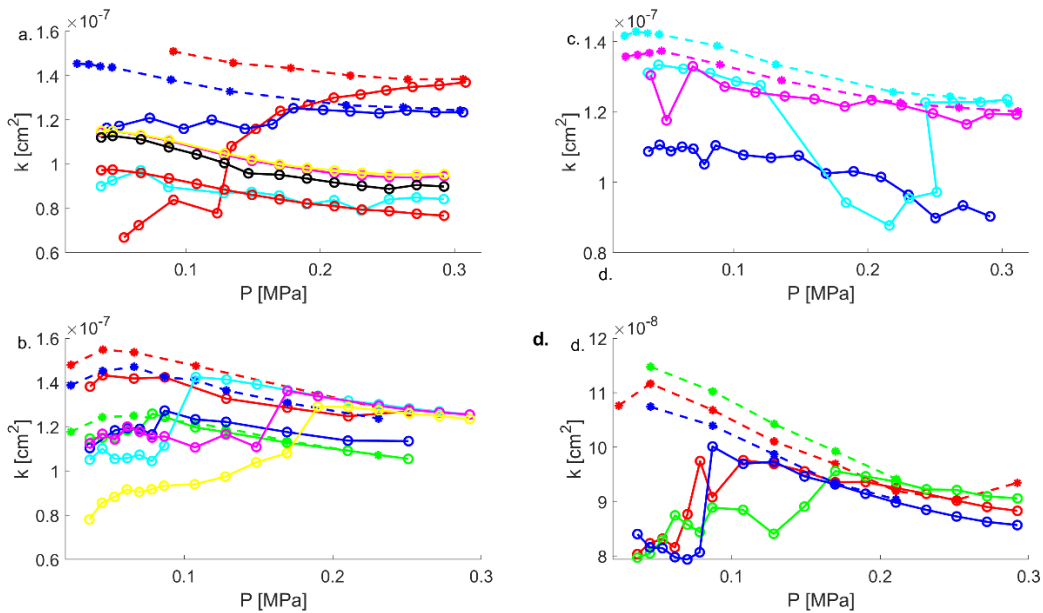

Figure 5. Permeability Vs applied pressure for concentrations of a. 5%, b. 7.5%, c. 10%, and d. 12.5%. For each concentration there are multiple repetitions done on different samples marked by their color, and as the sample becomes stiffer (higher concentrations), the repetitions have less variability. Circles mark a singular experiment, while asterisks at the same color mark a repetition of the same experiment on the same sample.

Utilizing the measured flux and pressure, in conjunction with the known porosity, fluid viscosity, and sample dimensions, the permeability for each applied pressure and its temporal evolution is computed using the radial Darcy law (equation 5). For the purposes of this study, the mean permeability calculated for the final 2 seconds preceding pressure termination is employed, where both the pressure and flux are stable, thereby yielding the steady-state permeability. In accordance with the derivation delineated in section 2.3, any alteration in permeability is ascribed to a change in porosity, which adheres to the role of effective stress. Consequently, the permeability change serves as a proxy for the strain change [5,66]. The steady-state permeability is plotted against the applied inlet pressure, functioning as a proxy for a stress-strain curve, where the permeability represents the strain, and the applied pressure symbolizes the stress (refer to figure 5). This approach aligns with previous work on elastic deformation [11]. For each sample, pressure is applied at the inlet for a specified duration while monitoring the pressure and flux. Subsequently, the applied pressure is halted, and the experiment is repeated with an elevated pressure.

3.2. The permeability response to the transition between ductile and brittle deformation

As stated in the previous section, the steady-state permeability can be calculated from the measured inlet pressure and flux, averaged over the entire sample, utilizing the radial Darcy law as expressed in equation 5. However, it is essential to recognize that the permeability changes as the inlet pressure escalates. In Figure 5, the permeability is plotted against the applied inlet pressure for various concentrations, with different colors representing distinct samples. A notable observation is the substantial variability in the initial permeability value among samples prepared with identical protocols. This variability, however, diminishes with an increase in solvent concentration. The permeability predominantly exhibits a decreasing trend, but abrupt jumps are observed, corresponding to the brittle deformations mentioned earlier. These deformations tend to occur at lower pressures as the solvent concentration rises. In some samples, the experiment is repeated to illustrate the effect of deformation on permeability, marked by an asterisk with the same color as the initial experiment. It is observed that the initial permeability commences at a higher pressure, followed by a gradual decrease until the values of both the initial experiment and repetition align. A similar observation was recently made by Hu et al [35] where sandstone subjected to increasing stresses during hydrostatic and triaxial compression exhibited either an increase in permeability due to microfractures or a decrease in permeability owing to microfracture sealing between loading and unloading cases. In the context of this study, the decrease in permeability is attributed to the ductile compression of the porous media skeleton in the plastic regime, followed by local or continuous brittle fractures that locally augment the permeability. As the compressive ductile phase intensifies with applied pressure, the permeability further diminishes as the fractures are compressed. However, when the experiment is repeated on the deformed sample, the newly formed fractures serve as conduits at low pressures, enhancing the permeability. As the pressure is further elevated, the same fracture compression transpires, and the conductivity aligns with the initial values from the preceding experiment. This process is contingent on the material strength of the sample, and on average, the overlapping permeabilities manifest earlier as the solvent concentration and initial porosity increase. Conversely, for low solvent concentrations, the variability among samples amplifies due to heightened sensitivity to air moisture and temperature during the evaporation process.

3.3. Modeling and constraining the effect of ductile to brittle deformation on permeability.

In the conducted experiments, the pressure difference was incrementally increased between each injection cycle, enabling the differentiation between the ductile and brittle regimes through permeability calculations. In the ductile regime, the material response was characterized by a reduction in permeability, while in the brittle regime, an abrupt increase in permeability was observed. According to established theory, ductile deformation leads to radial compression deformation, consequently reducing permeability in a spatially non-uniform manner. This non-uniformity in the effective stress, stemming from the combination of radial pressure drop and radial strain, results in peak compression between the sample boundaries. Such behavior is distinct from the compression deformation seen in 1D at the outlet boundary, a deviation attributable to the contrast between stress uniformity and linear pressure drop in the 1D, and the stress radial cubed drop and pressure radial drop for the circular case.

The hypothesis of this study suggests that this maximal radial compression triggers an instability that alters the angular stress alignment from parallel to the radial stress to a perpendicular orientation, specifically at the point where brittle fracture occurs. This transition in stress alignment can be interpreted as either the sum or the difference of the stresses, as detailed in section 2.3.5. To substantiate this hypothesis, the elastic and elasto-plastic stress-strain curves obtained from the pulling test (section 2.2) were utilized to derive the functional forms for both elastic (linear) and elasto-plastic (power-law) material responses, as elaborated in sections 2.3.2 and 2.3.4, respectively. This approach provided a comprehensive understanding of the material's behavior under varying stress conditions, contributing valuable insights into the complex interplay between permeability, deformation, and material properties.

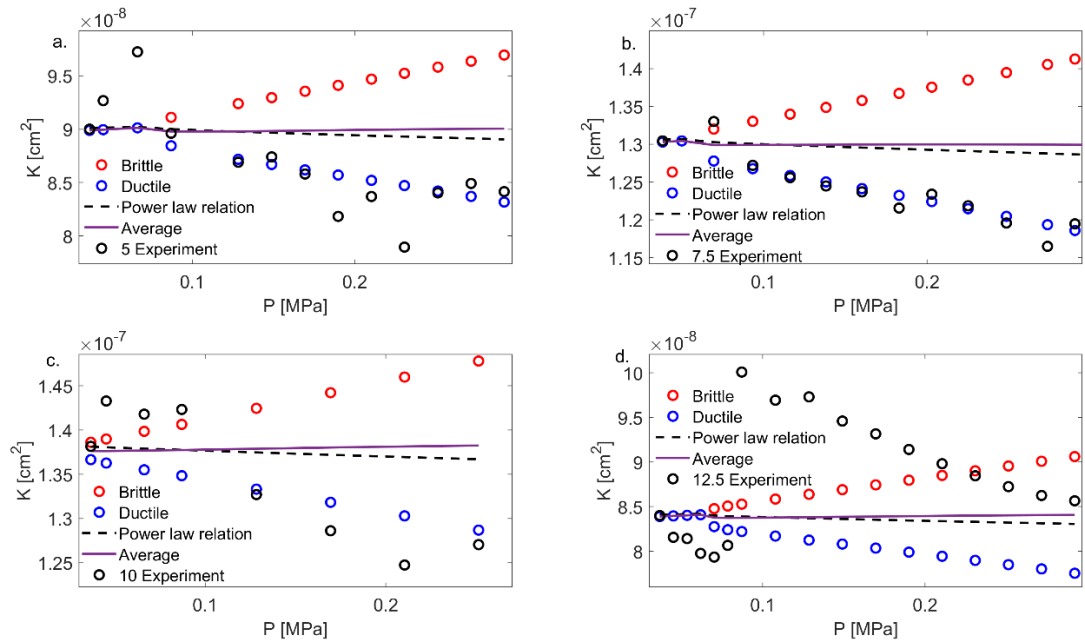

Figure 6. Permeability Vs applied pressure for concentrations of a. 5%, b. 7.5%, c. 10%, and d. 12.5%. The black circles are the measured permeability for the radial experimental setup, and the blue and red circles are the permeability for the ductile, and brittle cases respectively, calculated by the simplistic effective stress presented in section 2.3.1. The solid line is the averaging of the ductile and brittle case, while the brackish line is the permeability calculated by the power-law direct calculation in equation (22).

The identification of the threshold effective stress, where deformation transitions from elastic to plastic, facilitated the solution of the elastic or elasto-plastic poro-mechanic equation for radial and angular components. Utilizing the known effective stress for each radius (as detailed in sections 2.3.1 and 2.3.3), the radial and angular strains were determined. These strains were then multiplied by the porosity, based on the assumption that the volumetric strain ratio aligns with the porosity strain ratio. Consequently, for each local change in porosity, the corresponding change in permeability was calculated, and the results were radially averaged to ascertain the total permeability change due to pressurized flow (as described in section 2.3.5). This process was repeated twice: once considering the additive effects of radial and angular strain on permeability, and once considering the difference between angular and radial strains. Upon comparison of the measured permeability with the calculated permeability (illustrated in Figure 6 for the simplistic case), it was observed that the deformation corresponds with the ductile deformation, as indicated by the sum of angular and radial strains.

However, with the onset of brittle deformation, the increase in measured permeability was found to be confined by the permeability calculated from the difference between angular and radial strains. Notably, the measured permeability paralleled the trend and value for the ductile permeabilities, while the permeability calculated from the direct approach (equation 22) did not capture the decrease in permeabilities. This alignment between calculated and measured values, while confining the brittle response, without any fitting parameters, underscores the validity of the approach and provides a nuanced understanding of the interplay between ductile and brittle deformation, permeability, and porosity in the studied material.

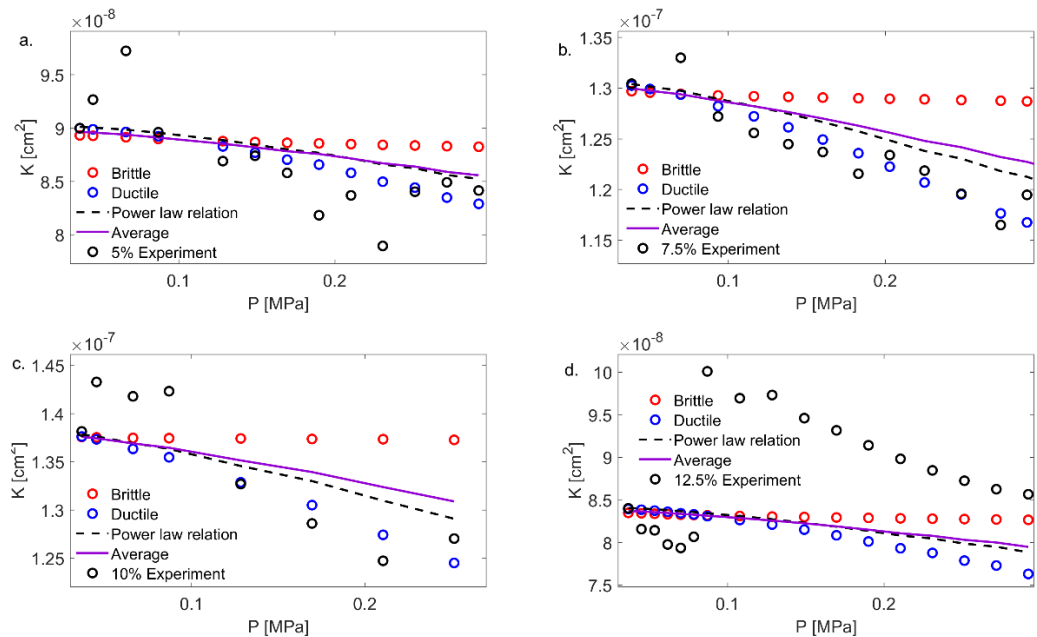

Figure 7. Permeability Vs applied pressure for concentrations of a. 5%, b. 7.5%, c. 10%, and d. 12.5%. The black circles are the measured permeability for the radial experimental setup, and the blue and red circles are the permeability for the ductile, and brittle cases respectively, calculated by the full effective stress presented in section 2.3.2, and 2.3.3 for the elastic and plastic cased respectively. The solid line is the averaging of the ductile and brittle case, while the brackish line is the permeability calculated by the power-law direct calculation in equation (22).

The calculation was repeated using the effective stresses for the full case, as delineated by [2], and adjusted for the specific setup as detailed in sections 2.3.2 and 2.3.4, for the elastic and plastic cases, respectively (as depicted in Figure 7). For the more ductile cases, characterized by high porosity (Figure 7 b and c), the deformation aligns with the ductile calculation, capturing well the mean trend of permeability reduction. Conversely, for the lower porosities (Figure 7 a and d), which also commence with lower permeabilities, the model misses the more brittle response. This discrepancy is a result of the dictated ratio between the radial and angular stress (as expressed in equations 15.1 and 15.2), yet the model is better in from the tic one in the fact that we do not need to find where the transition between elastic to elasto-plastic occur by total effective stress.

Interestingly, the Ryshkewithch model, as presented in section 2.3.5, adeptly captures the transition between various concentrations and porosities. The shift from the elastic case, where the radial and angular strains are identical for the simplistic case (and very close for the full model), to the elasto-plastic case, where a distinction between ductile and brittle deformation occurs, parallels the rapid change in permeability. This congruence underscores the model's efficacy in representing the complex interplay between permeability, porosity, and deformation, particularly in the context of transitions between different mechanical behaviors.

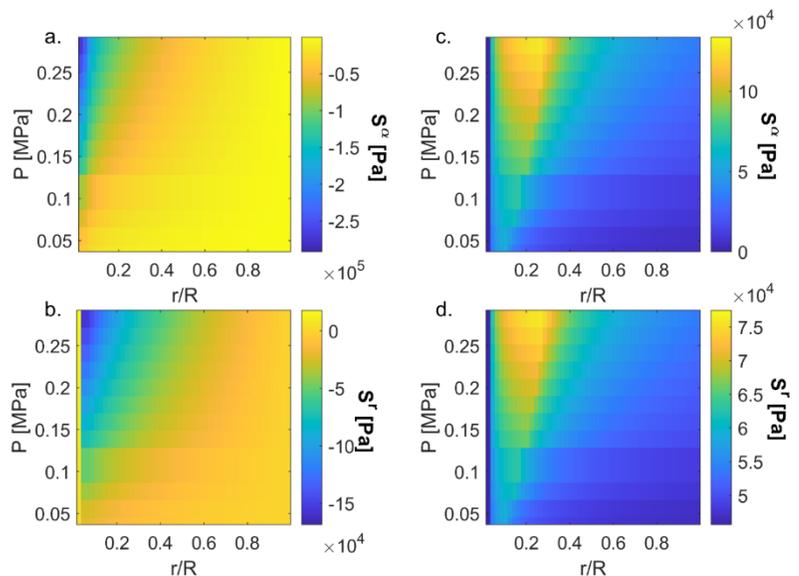

Figure 8. Effective stress calculation for the 5% case, as a function of applied inlet pressure and normalized radii for the a. tangential and b. radial stress for the simplistic calculation, and the c. tangential and d. radial stress for the full calculation case.

3.4. Comparing the redial strain effect on the ductile to brittle deformation and permeability.

The non-monotonic nature of the effective stress in radial pressurized flow has been noted in numerous studies, and it is indeed observed in the examination of the effective stress for both calculations in this context. There is a marked non-monotonic change in the effective tangential and radial stress, as depicted in Figure 8. For the simplistic case (Figure 8 a. and b.), the change is gradual, whereas the non-monotonicity for the full case is far more pronounced (Figure 8 c. and d.). Despite these differences in behavior, both models yield similar results for the overall permeability change and report a comparable range for the effective stress. The extent of the effective stress increases with the rise in inlet pressure, as anticipated, and encompasses a broader portion of the radii. This observation further underscores the complex interplay between pressure, stress, and permeability in radial flow systems, and the importance of accurately modeling these interactions to understand and predict the behavior of porous materials under varying conditions.

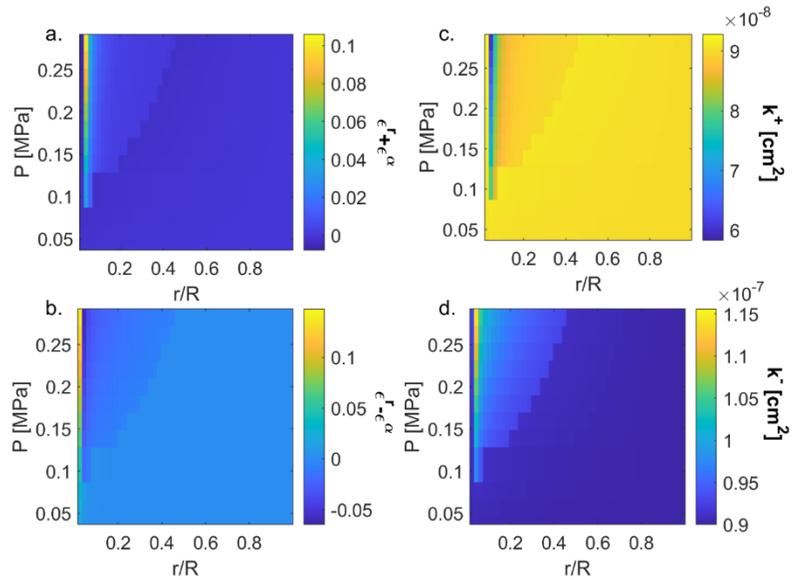

Figure 9. The simplistic calculation of the a. summation and b. subtraction of the radial and tangential elasto-plastic strains for the 5% case, as a

The non-monotonic nature of the effective stress leads to a corresponding non-monotonic strain, which is manifested in the transition between the regimes of plastic deformation and the elastic regime. This complex behavior is illustrated in Figures 9 and 10, where the total strain for both ductile and brittle cases is plotted against the inlet pressure and normalized radii. For the simplistic calculation, as the pressure increases, a distinct region of large plastic deformation emerges, while outward from this region, the deformation reverts to being small and elastic. This plastic region expands with the increase in inlet pressure for both ductile and brittle cases, mirroring the pattern of effective stress in both forms of calculation. In the full model approach, the rise and fall of the strain are more gradual for the ductile case, likely due to the continuity between the plastic and elastic regions with their corresponding radii. In contrast, the ductile strain for the simplistic case is more abrupt, stemming from the local criteria that define the transition between plastic and elastic responses using effective stress.

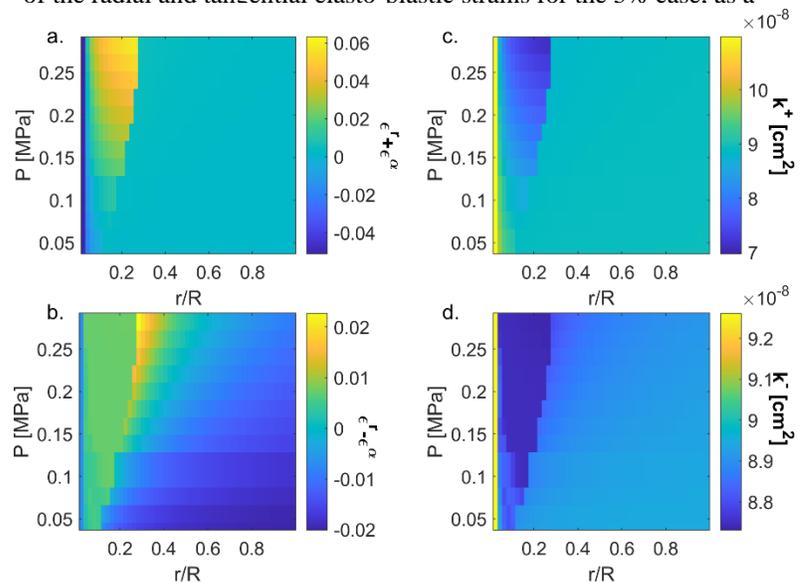

Figure 10. The full calculation of the a. summation and b. subtraction of the radial and tangential elasto-plastic strains for the 5% case, as a function of applied inlet pressure and normalized radii, and their corresponding permeability calculation, c. $k^+$, and b. $k^-$, respectively.

When examining the brittle case, both calculations reveal a sudden change in the plastic regime, marking the abrupt alteration in fracture formation. This behavior aligns with the concept of brittle fracture, which advances rapidly due to local low pore pressure in the newly formed conductive path and the relatively high effective stress, leading to continuous or semi-continuous fractures. The total brittle deformation is confined by the brittle calculation, but instead of occurring abruptly, it accumulates gradually. Following the brittle deformation, ductile compressing deformation ensues, reducing the overall permeability. Thus, both dilation of the brittle fracture and compression of the ductile deformation occur simultaneously within the sample. Furthermore, both brittle and ductile effects on permeability are confined within the simplistic and full calculation without any fitting parameters, accommodating various initial permeabilities and sample strengths. A closer examination of the strain reveals a marked difference between the simplistic and full calculations, with the simplistic calculation exhibiting higher strains. This discrepancy can be traced to the difference in stress decrease, which is inversely proportional to the radii for the full calculation, compared to the inverse cube of the radii for the simplistic calculation.

4. Summary and conclusions

This study systematically investigated the coupling between permeability change and deformation in PMMA rock-like porous material, with a specific focus on the transition between ductile and brittle deformation. Experimentally, the research involved the application of incremental pressure differences across various injection cycles, enabling the distinction between ductile and brittle regimes. The material's response was monitored through permeability calculations, revealing a non-uniform reduction in permeability during the ductile regime and abrupt increases during the brittle regime. The experimental setup allowed for the reproduction of saturated pressurized flow, applied radially for four different material strength group, mimicking well injection scenarios.

To capture the elastic and plastic response for both the ductile and brittle cases, the study employed both simplistic and full calculations to simulate the material's behavior. The modeling involved the derivation of functional forms from the pulling tests for the elastic (linear) and elasto-plastic (power-law) material responses, using the stress-strain curves. The threshold effective stress was identified, marking the transition from elastic to plastic deformation. The elastic or elasto-plastic poro-mechanic equation was solved for radial and angular components using the known effective stress for each radius. This enabled the determination of radial and angular strains, which were then related to porosity changes. The permeability change was calculated for each local porosity change and averaged radially to ascertain the total permeability change due to pressurized flow.

Two approaches were considered for the calculation of permeability: one adding the effects of radial and angular strain to capture the ductile deformation and the other subtracting the angular strain from the radial strain to capture the brittle deformation. While the ductile deformation capture the permeability decrease, the brittle confined the permeability increase. The difference between the ductile and brittle fracture can also be ascertain from figure 2, where the ductile response to the stress has a long elongation response compared to the elastic regime, in contrast to brittle response that occurs more suddenly. However, due to the effective stress radial change, both ductile and brittle deformations occur at different regions in the sample, as the angular strain can be added or subtracted from the radial strain, as proposed in our study.

Both the simplistic and full calculations were found to capture the material's response, with the simplistic calculation confining the brittle response better, while the full calculation provided the transition radii between elastic to plastic. The modeling also showed the known non-monotonic change in the effective tangential and radial effective stress, leading to a corresponding non-monotonic strain. This was manifested in the transition between the regimes of plastic and elastic deformation. The study also utilized the Ryshkewithch model, which captured well the transition between various concentrations and porosities.

In conclusion, the research has unveiled a complex interplay between stress, strain, and permeability in porous materials. The non-monotonic nature of effective stress led to corresponding non-monotonic strains, with distinct regions of plastic and elastic deformation. The study successfully captured both ductile and brittle deformations, revealing how they occur simultaneously within the sample.

The findings demonstrate that the simplistic and full calculations can confine both brittle and ductile effects on permeability without any fitting parameters, accommodating various initial permeabilities and sample strengths. The study also highlighted the marked difference between the two calculations, with the simplistic approach exhibiting better confinement for the brittle deformation and the full approach providing the transition radii for the plastic to elastic case directly.

Moreover, the research has shown that the transition between the elastic case to the elasto-plastic case is akin to the rapid change in permeability, providing a nuanced understanding of the material's behavior. The insights into the abrupt nature of brittle fractures and the gradual accumulation of deformation contribute to a comprehensive understanding of the material's response.

This study not only advances the scientific understanding of porous materials but also offers valuable insights for practical applications, particularly in the context of sedimentary rocks and hydraulic deformation, which was shown by Hu et. al, [35]. The findings underscore the importance of accurate modeling and the consideration of both ductile and brittle deformations in predicting and interpreting the responses of such materials. The methodologies and insights derived from this research hold significant potential for further exploration and application in related fields.

Acknowledgments
Y.E., S.S. and M.S. thank the support of ISF-NSFC (grant No. 3333/19).

Supplementary to Investigating the Permeability Evolution of Artificial Rock During Ductile and Brittle Deformation Under Pressurized Flow

Shaimaa Sulieman[1], Martin Stolar[1], Ludmila Abezgauz[1], Shouceng Tian[2], and Yaniv Edery[1]

[1]Faculty of Civil and Environmental Engineering, Technion, Haifa, Israel.

[2]State Key Laboratory of Petroleum Resources and Prospecting, China University of Petroleum (Beijing), Beijing, 102249, China

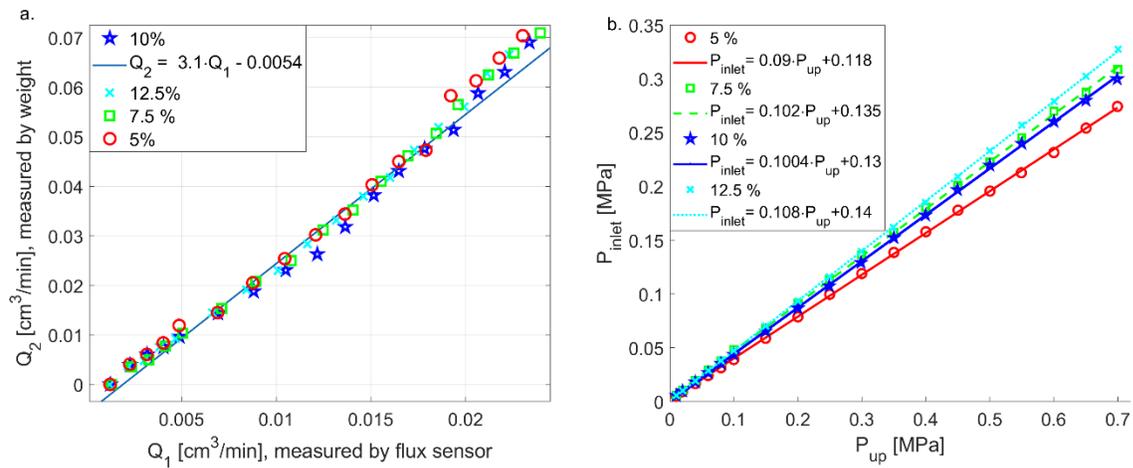

Figure S1. a. Calibration curve from the flux sensor ($Q_1$) to the actual flux measured by analytical weight for each applied pressure ($Q_2$), and for each sample preparation. b. Calibration curve from the applied pressure ($P_{up}$) to the measured pressure at the inlet ($P_{inlet}$), measured by a pressure sensor for each applied pressure, and for each sample preparation.

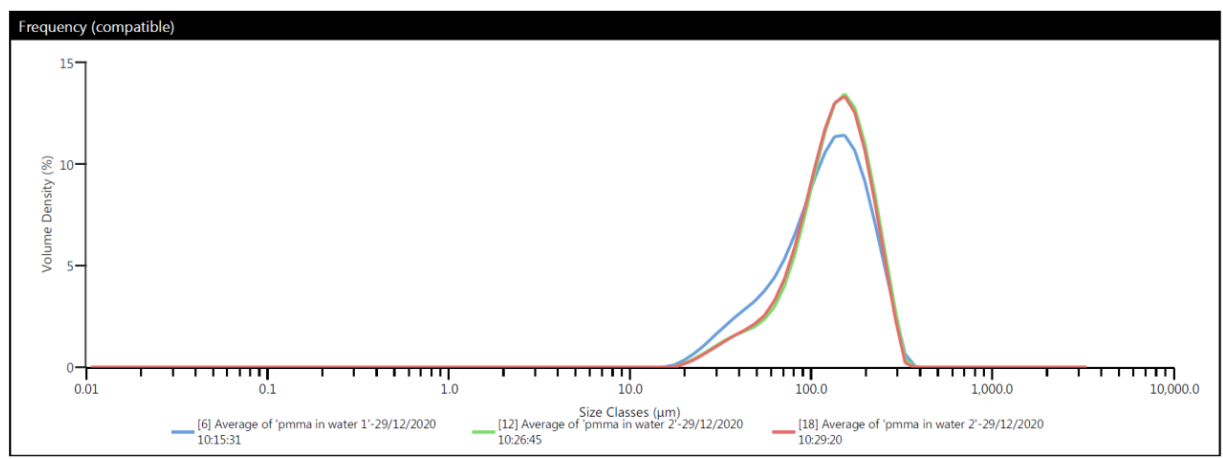

Figure S2. PMMA bead size distribution measured by MasterSizer (Three repetition).